\pdfoutput=1
\documentclass{article}[12pt]

\usepackage{graphicx}
\title{An evaluation of a microprocessor with two independent hardware execution threads coupled through a shared cache}
\author{Madhav P. Desai\\ Department of Electrical Engineering\\ IIT Bombay\\ Mumbai, India}
\begin{document}

\maketitle
\begin{abstract}

We investigate the utility of augmenting a microprocessor with a 
single execution pipeline by adding a second copy of the 
execution pipeline in parallel with the existing one.
The resulting dual-hardware-threaded microprocessor has
two identical, independent, single-issue in-order execution 
pipelines (hardware threads) 
which share a common memory sub-system (consisting of instruction and data 
caches together with a memory management unit).
From a design perspective, the assembly and verification
of the dual threaded processor is simplified by
the use of existing verified implementations of the 
execution pipeline and a memory unit.   Because the memory unit is 
shared by the two hardware threads, the relative area overhead of 
adding the second hardware thread is 25\% of the area of
the existing single threaded processor.

Among the benefits of the shared memory subsystem
is the tight coupling of the two execution threads
via the shared level-1 caches, making it easy for them to share
information. Among the drawbacks is the destructive interference
between the two threads in terms of stalls due to cache misses.
In order to understand the impact of
the benefits and drawbacks, we use FPGA implementations to
compare the dual threaded processor with
the original single threaded processor on a set of representative
applications.

We use a lock-less cooperative
programming model for the mapping of applications 
to the dual-threaded processor
system while maintaining compatibility with existing software.
Over a wide range of applications with internal parallelism, 
we observe that performance on 
the dual-threaded processor is
substantially better than on the single-threaded version.
In applications without internal parallelism,
the speedup is not significant, as expected.  Also, the dual-threaded
performance  can be degraded when one of the threads 
causes a large number of cache misses or generates 
an excessive number of loads and/or stores.

\end{abstract}

\section{Introduction}

The capacity of a micro-processor can be increased chiefly in three ways: 
by increasing the number of instructions that the processor issues every
clock cycle (super-scalar processors), by increasing the number of hardware threads 
in the processor (multi-threading), and by increasing the number of cores
in the processor (multi-core).

We investigate the second approach.  Our starting point
is a processor having a single execution thread, namely,
the AJIT processor (see Section \ref{sec:AjitIntro}).  In this
processor, we observe that the execution thread fetches a pair of
instructions per access and uses approximately half
the available bandwidth of the instruction cache.  Further,
the fraction of load/store instructions in
a wide range of single threaded applications \cite{ref:InstrMix}
is less than 50\%.   Thus, the caches can potentially serve
two execution threads simultaneously without becoming a bottleneck.

Starting with the implementation of
a micro-processor with a single execution pipeline, we augment it
by adding a second, identical execution thread (pipeline) so that the
two threads work with a single memory unit, which
consists of an instruction cache, a data cache, and a memory management unit.
The two parallel execution threads are closely coupled via the caches in
the shared memory unit.  It is possible to de-activate one of the
threads so that power consumption is reduced when that thread
is idle.

The cost of the dual-threaded processor is much lower than
the cost of a {\em dual-core} processor due to the sharing of the caches and memory management
unit.  Based on data from a 65nm implementation \cite{ref:IITBNAVIC} of the pre-existing 
single threaded processor with 32KB of instruction and data cache, with a memory management unit 
(2.5 sq.mm for the entire processor, out of which 75\% is due to the caches and
the memory management unit), 
we observe that the area of the dual threaded processor will be 
1.25X larger than that of the single threaded processor.

In principle, if the shared memory unit does not become
a bottleneck, and if the workload is parallelizable, the dual-threaded 
processor should offer double the performance of a single-threaded
processor.  In Section \ref{sec:DualThreaded}, we provide
a justification of why we do not expect the shared memory
to become a bottleneck in typical applications.   This is
corroborated by the performance increase
observed across several
applications ranging from sorting to fast-fourier-transforms
and a complete electro-cardiogram signal processing chain.

On the other hand, the dual-threaded processor has some disadvantages.
The effective available memory bandwidth is shared by the
two hardware threads.   Thus, applications which are memory intensive may
not exhibit the expected speed-up.  The instruction and data access patterns generated 
by the two threads may induce mutual cache misses and cache thrashing.  
Also, locking of the memory unit by one of the threads (for example, on an atomic instruction) 
can stall the execution of the other thread.
Slow I/O memory accesses by one thread can block memory accesses 
by the other thread.

In our evaluation, we implement the single-threaded and dual-threaded
processors on a common FPGA platform and measure the
relative performance.   A cooperative programming model is
used to parallelize applications across the two threads.
The following questions are addressed:
\begin{itemize}
\item  What is the time overhead in the synchronization of
activity across the two threads in the dual-threaded processor?  This
turns out to be 25 clock cycles for the FPGA implementation with
the cooperative programming model used in the characterization.  
\item How does the dual threaded processor compare with the single
threaded processor in terms of performance?   We look at three scenarios
in which the dual-threaded processor can be used; both threads working
on an application, one thread working and the other spinning for work,
and one thread working and the other thread de-activated.  
We measure the performance in each of these scenarios (as well
as on the single threaded processor)  on ten test applications. 
\end{itemize}

We observe significant ($ > 1.6X$)
speedup  in the dual threaded (both threads active and working) processor
relative to the single threaded reference processor in seven of the ten
applications.  In two of the applications, performance improvements
were lower, and in one of the applications, there was a degradation in
performance when using the dual threaded processor.
For performance critical algorithms which have
internal parallelism, the performance increase observed
due to the use of the second thread is far higher than
the extra cost incurred in adding the second thread to the processor.

\section{Other approaches to multi-threading reported in the literature}

A software thread is a sequence of instructions generated by 
a sequential program, and a hardware thread is a logic circuit
which executes instruction streams.   Multi-threading techniques
involve the mapping of multiple software threads onto one or more
hardware threads at run time.
A detailed survey of explicit multi-threading
approaches is given in \cite{ref:MultiThreading}.

The simplest multi-threading
approach involves the multiplexing of multiple software threads
onto a single hardware thread.   Whenever a
software thread stalls due to a slow memory access, the hardware
thread can switch to the execution of a different software thread.
The hardware thread may itself be a simple single issue pipeline or
a complex super-scalar pipeline.
At any point in time, the hardware thread may be executing
instructions from either a single software thread or multiple software
threads (simultaneous multi-threading).   

The mechanism used to select the software thread to execute can
be of many types. 
In blocked multi-threading, a software thread runs on the harware
thread until the software thread blocks (on a cache miss, for example), 
at which point the hardware thread switches to a different software thread.
Interleaved multi-threading is a variant in which 
multiple software threads are interleaved at the fetch stage
itself into a single stream which then executes on a hardware thread.

In our approach, there are two identical hardware threads that
are available for the execution of software threads.   These
threads are tightly coupled via the caches.   This is coarse grained
symmetric multi-threading \cite{ref:MultiThreading} with two parallel 
independent threads, in which the workload of the two hardware threads 
is managed explicitly by software.  The problem of data coherency across 
the two hardware threads does not arise because of the shared level-1
caches.

The proposed scheme is different from a multi-core implementation
because each core in a multi-core has its own cache
subsystem and special effort is needed to keep the caches coherent across
the cores.

\section{The starting point: the single-threaded AJIT processor} \label{sec:AjitIntro}

\begin{figure}
  \centering
  \includegraphics[width=10cm]{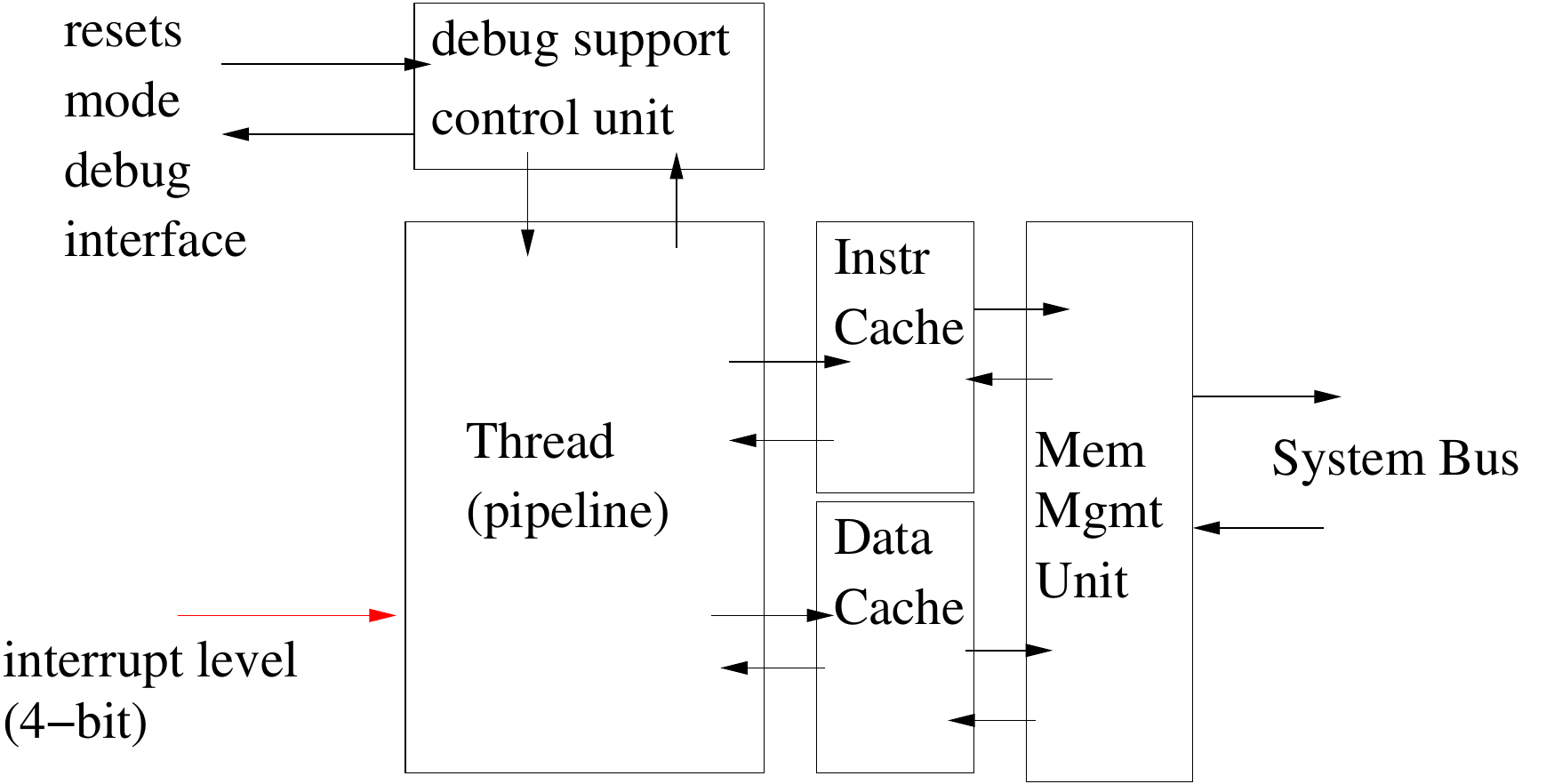}
  \caption{AJIT single-threaded core}
  \label{fig:AjitSingleThreadTop}
\end{figure}

We describe the starting point of our implementation of a dual-threaded processor.
The AJIT single-threaded processor developed at the Indian Institute of Technology (Bombay) is
an implementation of the SPARC V8 \cite{ref:V8} instruction set architecture.   The 
single-threaded processor has the structure shown in Figure \ref{fig:AjitSingleThreadTop}.

The principal components of the architecture shown in Figure \ref{fig:AjitSingleThreadTop}
are the execution pipeline (or thread),  the instruction and data caches, and
the memory management unit.   The control and debug unit is responsible for the
thread initialization, exception handling, and for providing a debug mechanism 
to a remote debugger.  

The single threaded AJIT processor boots Linux (3.16.1), has been
verified extensively, and has been used in the implementation of
a GNSS base-band receiver system-on-chip (in 65nm technology) \cite{ref:IITBNAVIC}. 
This system-on-chip provides base band processing needed for GPS and IRNSS based 
satellite positioning systems.

We will describe the principal components in brief.

\subsection{The execution pipeline (the thread)}

\begin{figure}
  \centering
  \includegraphics[width=12cm]{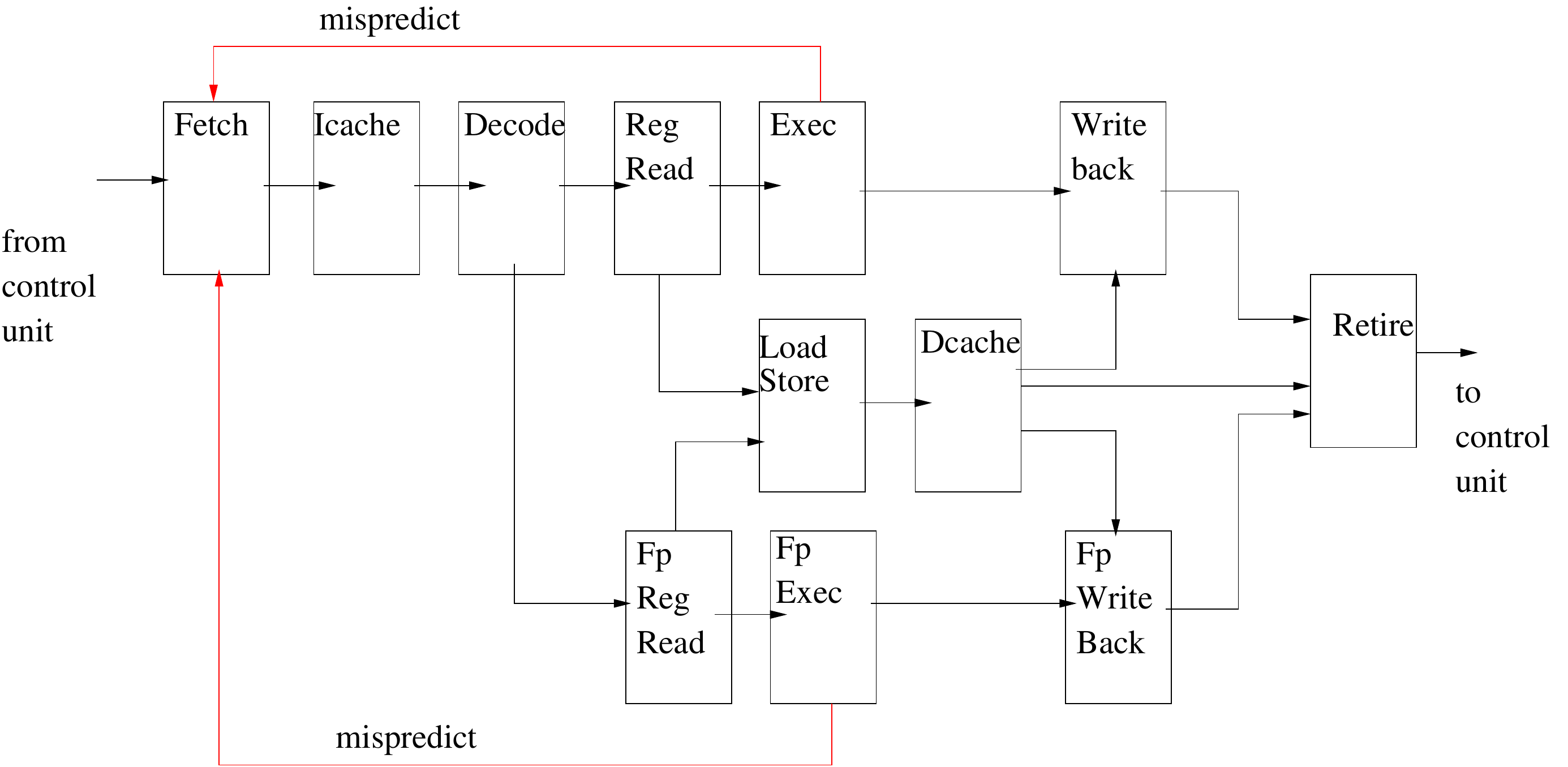}
  \caption{AJIT execution pipeline}
  \label{fig:AjitThread}
\end{figure}

The execution pipeline is shown in Figure \ref{fig:AjitThread}.
It is a single-issue, in-order, elastic pipeline.  Every arrow in
Figure \ref{fig:AjitThread} is physically implemented as a first-in-first-out queue (FIFO).
The depths of the queues are carefully chosen to maximize pipeline
performance across a variety of application programs.   Further,
the flow of an instruction down the pipeline depends on the kind
of instruction.  For example, an integer arithmetic operation
has the flow
\begin{verbatim}
Fetch -> Icache -> Decode -> Reg Read -> 
                 Exec -> Write Back -> Retire
\end{verbatim}
whereas, a load operation to the integer register file has the flow
\begin{verbatim}
Fetch -> Icache -> Decode -> Reg Read -> 
                Load store -> Dcache -> Write Back -> Retire
\end{verbatim}

The fetch unit includes a branch target buffer, a
2-bit branch predictor and a return address stack.   The integer execution unit
implements all arithmetic and logic instructions in a single clock cycle
(except for the divider).   The floating point unit implements single and
double precision operations.  Multiply and addition are pipelined two-cycle operations, 
while square-root and divide are non-pipelined and take 24 (resp. 16) clock cycles
for the double (resp. single) precision case.  All exceptions are handled in a precise
manner.

The resource utilization of a single execution thread, when implemented on 
a Xilinx Kintex xc7k325tffg900-2 FPGA, is shown in Figure \ref{fig:SingleThreadUtil}.
\begin{figure}
  \centering
\begin{verbatim}
------------------------------------------
 Unit                   LUTs    Flip-flops
------------------------------------------
fetch                   4.1K      3.2K 
decode                  1.4K      0.76K
registers (int)         5.8K      2.7K
execute (int)           4.4K      2.99K
loadstore               4.1K      3.2K
fpu                    20.7K     10.1K
debug/control unit      4.5K      5.3K
------------------------------------------
\end{verbatim}
	\caption{Resource utilization of a single thread on the Xilinx xc7k325tffg900-2 FPGA}
	\label{fig:SingleThreadUtil}
\end{figure}

\subsection{The memory unit}

The memory unit consists of virtually-indexed virtually-tagged (VIVT)
instruction and data caches, with a memory management unit (MMU) which conforms
to the SPARC V8 reference MMU specification \cite{ref:V8}.

The VIVT data and instruction caches can be configured to sizes ranging from
4KB to 32KB.  The associativity of the caches can be configured from 1 (direct mapped)
to 8.  The cache line size is 64 bytes for both caches.
The data cache uses a write-through-allocate policy, and both caches use a 
not-most-recently-used replacement policy.   A synonym detection scheme 
and an invalidation mechanism for enforcing cache coherence in a multi-core
setting is also implemented in the memory unit \cite{ref:SynonymCacheCoherence}.
On a hit, the instruction and data caches provide data to the requesting
hardware thread in one clock cycle.

The memory management unit has a 256-entry 8-way set associative translation buffer,
and implements the page table walk algorithm specified by the SPARC V8 reference MMU
specification \cite{ref:V8}.

\subsection{Baseline performance of the single-threaded AJIT processor}

The base-line performance of the single-threaded AJIT processor on standard
benchmarks is summarized in Figure \ref{fig:benchmarks}.
\begin{figure}
  \centering
\begin{verbatim}
-------------------------------------------------------
Benchmark          Performance          
-------------------------------------------------------
Dhrystone          2.4   DMIPS/MHz       
Coremark           2.42  Coremarks/MHz  
Linpack            0.27  MFLOPS/MHz    
Whetstone          0.4   MWIPS/MHz    
-------------------------------------------------------
\end{verbatim}
\caption{Single-thread processor benchmark performance}
\label{fig:benchmarks}
\end{figure}
The GCC 4.7.4 compiler was used to compile these benchmarks.   The processor
was configured with 32KB, 4-way associative data and instruction caches, and
included a memory management unit.   
For the evaluation, the processor was implemented in FPGA and operated at 80MHz.
The external memory was 1GB of DDR3 dynamic memory.

\section{Structure of the dual-threaded AJIT processor} \label{sec:DualThreaded}

\begin{figure}
  \centering
  \includegraphics[width=8cm]{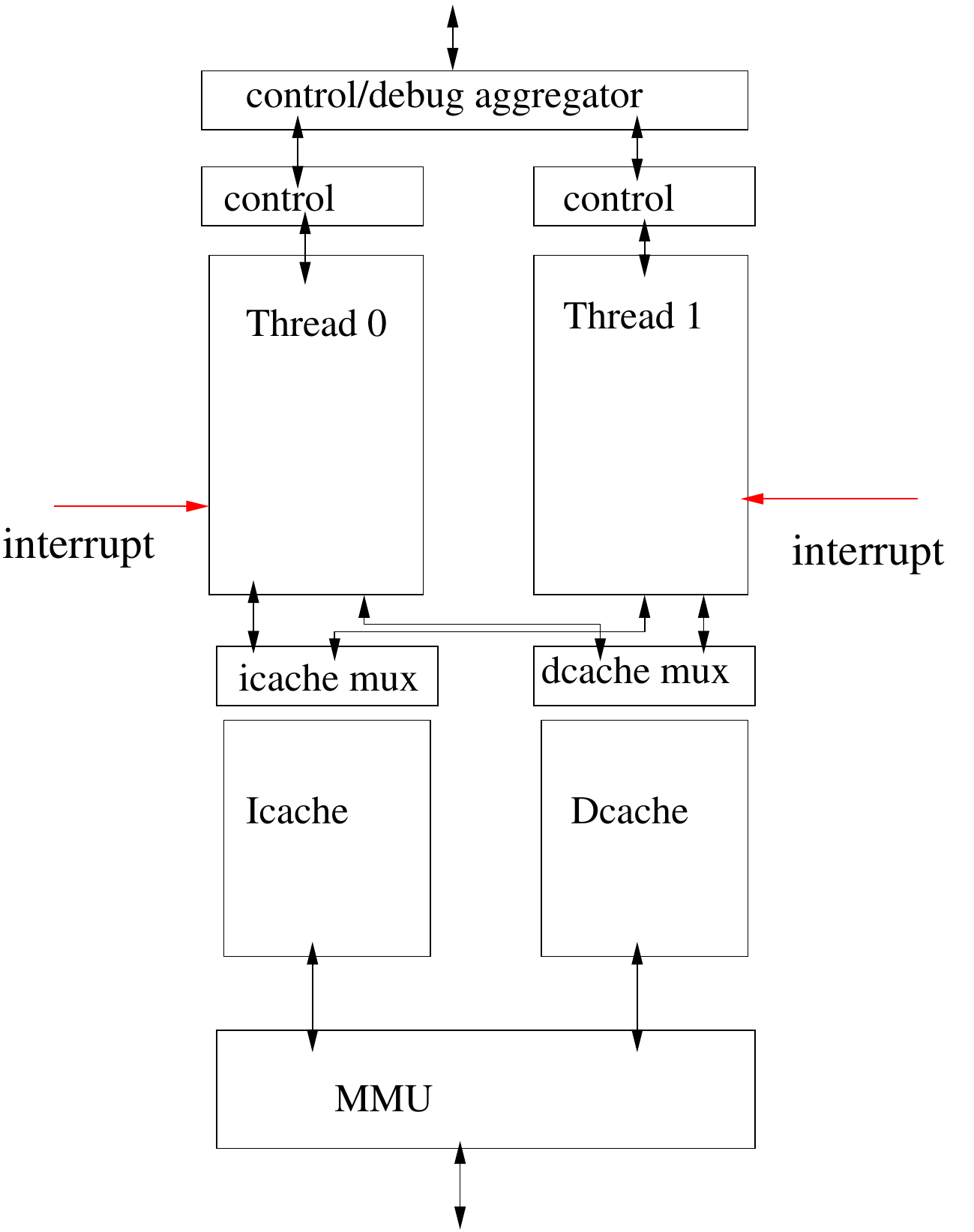}
  \caption{AJIT dual-threaded core}
  \label{fig:AjitDualThreadTop}
\end{figure}

The dual-threaded AJIT processor is constructed using two execution pipelines which
share the memory unit.  This is illustrated in Figure \ref{fig:AjitDualThreadTop}.
The two threads are identical in all respects and are connected to the
instruction and data caches with multiplexors.  Each of the execution pipelines
includes a small 128-entry instruction buffer in which recently 
accessed instructions are kept.  If there is a hit in this instruction buffer,
an instruction cache access is unnecessary.   This instruction buffer is useful
when the thread is executing a spin-loop.  

The cache multiplexors provide 
support for cache locking, and guarantee fair access to the caches for
both threads.   Further, it is possible to independently control
the two threads for remote debugging.  Since the internals of the two threads are 
not modified in any manner (except for the instruction buffer mentioned above), 
the verification effort for the dual threaded processor is considerably reduced. 

The interaction between the two threads in this processor is through
the caches, and is fast.  The latency in a message from one thread to
another can be as small as one clock cycle (a store by one thread is
visible to a load one cycle later).  This allows low latency synchronization
across the two threads and provides benefits for parallelizing work loads.

Further, we expect the performance of the dual threaded processor
to be superior to the single threaded processor because the caches are
unlikely to become a bottleneck for most programs running on the
two threads.
\begin{itemize}
\item An individual thread fetches two instructions on every access to the
instruction cache.  Thus, a single thread typically accesses the instruction
cache once every two cycles, and utilizes approximately 50\% of the instruction
cache bandwidth.   Thus, when two threads are present, the
instruction cache has sufficient bandwidth to serve both of them.
\begin{itemize}
\item In addition, each thread in the dual threaded AJIT processor includes
a small (128-entry) instruction buffer which caches recently accessed
instructions.  This reduces the pressure on the instruction
cache whenever one of the threads is executing a tight
spin-loop.
\end{itemize}
\item For most applications, the fraction of instructions that access
the data cache (load/store instructions) is less than 50\%.
For example, in the AJIT single threaded processor, the fraction of
load/store instructions for some of the common processor benchmarks
is shown in Figure \ref{fig:ls}.
\begin{figure}
  \centering
\begin{verbatim}
---------------------------------------------------
Benchmark        fraction of load/store
                  instructions 
---------------------------------------------------
Coremark          19%
Dhrystone         15.4% 
Linpack           31%
---------------------------------------------------
\end{verbatim}
\caption{Fraction of load/store instructions in benchmarks}
\label{fig:ls}
\end{figure}
An exhaustive analysis of the instruction mix across 
the entire set of SPEC (integer and floating point) benchmarks for the SPARC-V8 processor
has been reported in \cite{ref:InstrMix}.  In
all the SPEC integer benchmarks that they studied, the fraction of load/store
instructions was  never greater than 33\% (ranging from 17\% to 33\%).  
Across the SPEC floating point benchmarks, the peak fraction of load/store instructions was 
never greater than 47\% (ranging from 25\% to 47\%).   We see that for the
SPEC workload, the utilization
of the caches by a single thread does not exceed 50\%.  For
typical programs, sharing the
available cache bandwidth across two threads will not create a 
bottleneck.  
\end{itemize}

Note that even if the utilization of the caches by a single thread is
less than 50\%, it is possible that the performance of the dual thread
processor suffers relative to the single thread case due to cache misses
and cache contention.   The caches block on a miss, and thus,
one of the threads may be forced to stall due to a miss generated by the
other thread.   To mitigate this effect, we use large 32KB level 1 caches
with 4-way set associativity in order to increase the cache hit rate.

\section{Programming models used for the evaluation of the dual-threaded AJIT processor}

The software which runs on the dual-threaded core should ideally
be structured so that:
\begin{itemize}
\item The software offers a well understood, simple view of the
dual threaded processor.  This should be true even for bare metal
applications.
\item When both threads are active, they should be working on the same context, 
using the same page table.  This restriction is imposed by the use of 
VIVT caches.
\item To get performance improvements, the cache resident part of the
data used by the two threads must overlap as much as possible.
\end{itemize}

\begin{figure}
  \centering
  \includegraphics[width=8cm]{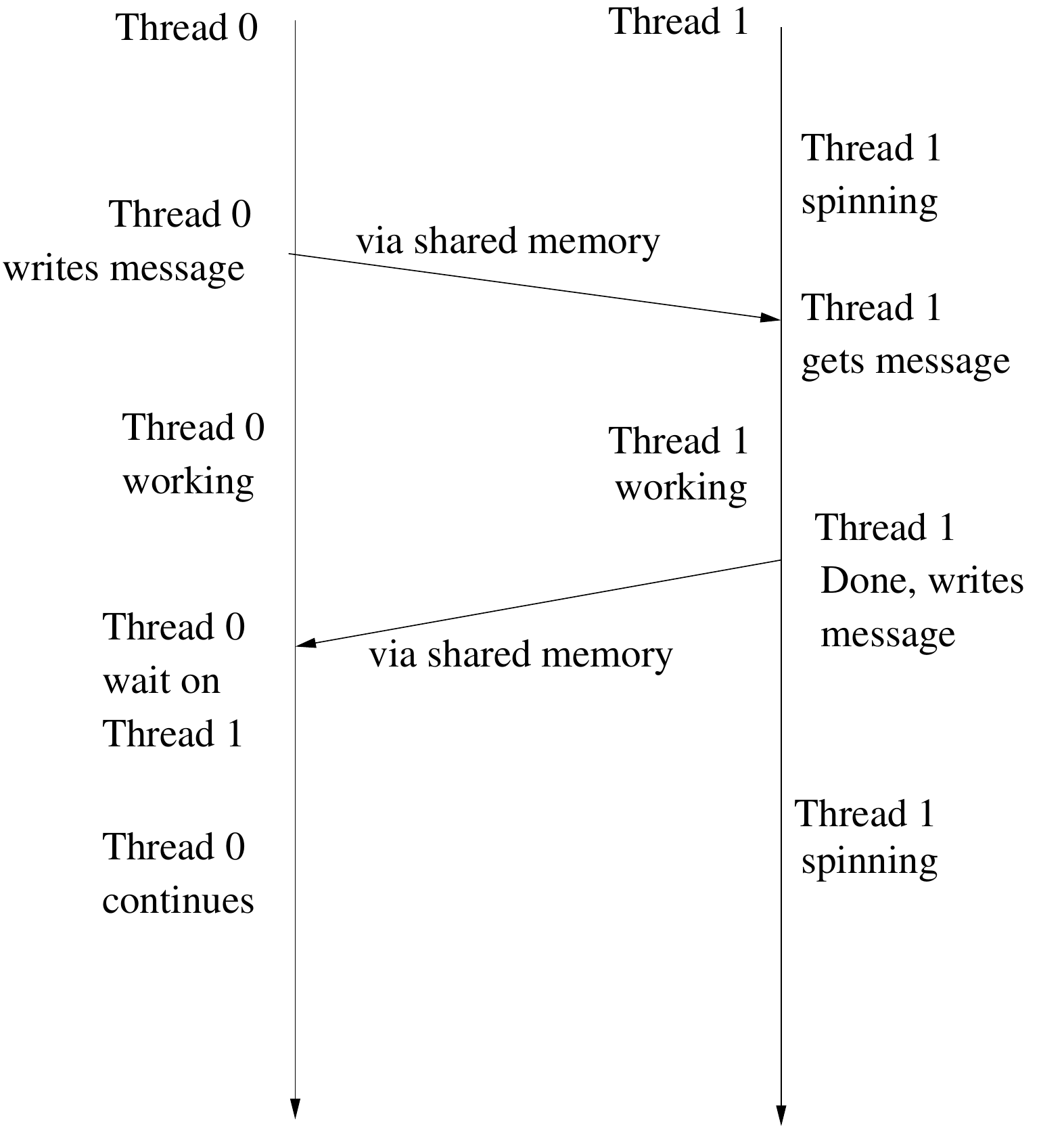}
  \caption{Event diagram in side-kick programming model}
  \label{fig:EventDiagram}
\end{figure}

For the current performance evaluation, we have proposed a programming model
that satisfies these requirements.  We call this the side-kick programming model.
Thread 0 is the main execution thread, and it uses thread 1 as a compute assistant.
When the assistance of thread 1 is required, thread 0 passes run-time information
to thread 1 via a shared memory buffer, which we refer to as a thread channel.  
Thread 0 effectively specifies a function and its arguments
to thread 1.   When thread 1 has finished executing the function, it indicates
the status, and returns values through the thread channel.
Note that while thread 1 is working, so can thread 0.   
The flow of events in the two threads can be visualized as shown in Figure \ref{fig:EventDiagram}.
This simple cooperation model can be implemented
without using locks.  The small instruction buffer in a thread 
will provide the instructions needed by that thread whenever it
is executing a tight spin loop (for example, waiting
for a message from the other thread).

The side-kick programming model is also easy to fit into 
existing software infrastructure.  The overall application runs
on thread 0, and only critical routines need to be re-written so that
they use thread 1 as an assistant.  The effort required to re-write the critical
routines is similar to that of writing a multi-threaded program
using {\em pthreads}.   

\subsection{Overhead for invoking a side-kick routine}

To understand the overhead for the invocation of a
task on thread 1 by thread 0,  we invoke a trivial task
on thread 1 from thread 0 and measure the round-trip
latency (invocation to getting return notification) on
thread 0.  This measurement is done on the
evaluation system described in Figure \ref{fig:EvalSystem}.
We observe that the round trip overhead latency is
$25$ clock cycles.   Assuming that a 25\% invocation overhead is
acceptable in practice, this implies that the granularity of the task that is invoked 
on thread 0 can be as small as $100$ clock cycles when partitioning work across
the two threads.

\section{Applications used to evaluate the dual-threaded processor}

The following application programs were used to characterize the
dual-threaded processor.
\begin{itemize}
\item Matrix-mult: two 128x128 matrices $A$ and $B$ 
with double precision floating point entries are multiplied using the
two threads.   Thread 1 handles the even rows of the  matrix $A$
and Thread 0 handles the odd rows.  
The computation is repeated 1024 times
during the measurement.
\item Dot-product:  a dot-product of two  1024 element vectors (using double
precision arithmetic).   Thread 1 calculates and sums the even products,
and Thread 0 does the same for the odd products.
The computation is repeated 1024 times
during the measurement.
\item FFT; a 4096 point fast fourier transform (using double precision floating
point values) was evaluated by the two threads.  The 4096 point FFT is
evaluated using two parallel 2048 point FFT's on the two threads.
\item Merge-sort: A 1024 element vector of integers was sorted using a
merge-sort algorithm using the two threads.  Half the numbers are sorted by
thread 1, the other half by thread 0, and the final merge is done by
thread 0.
The computation is repeated 1024 times
during the measurement.
\item Bellman-Ford: Given a graph, calculate the shortest paths between all
pairs of vertices in the graph.  Thread 0 is used to calculate 
the paths for half the nodes, and thread 1 is used for the other
half. A graph with 64 nodes and 128 edges is used,
The computation is repeated 1024 times
during the measurement.
\item Daxpy: Given a scalar $a$, and two 1024-element double precision vectors $X$ and $Y$, compute
\begin{displaymath}
aX \ + \ Y
\end{displaymath}
This is parallelized across the two threads by using thread 0 to update the
first half of the result, and thread 1 to update the second half.
The computation is repeated 1024 times
during the measurement.
\item Mem-copy: a 32KB block of memory was copied from one memory region to another,
using the two threads. Odd entries are copied by thread 1, even entries by
thread 0.
\item Mutexes: an integer is incremented under a mutex by the active threads.
The time required for the integer to reach a value of $2^{20}$ is measured.
\item An ECG signal processing pipeline, which we will describe in detail
later, in Section \ref{sec:ECG}.  Sixteen heart-beats are analysed during the
measurement.
\item Coremark: two copies of the coremark benchmark are run 
simultaneously on the two execution threads.  
\end{itemize}

\section{Performance} \label{sec:Performance}

\begin{figure}
  \centering
  \includegraphics[width=8cm]{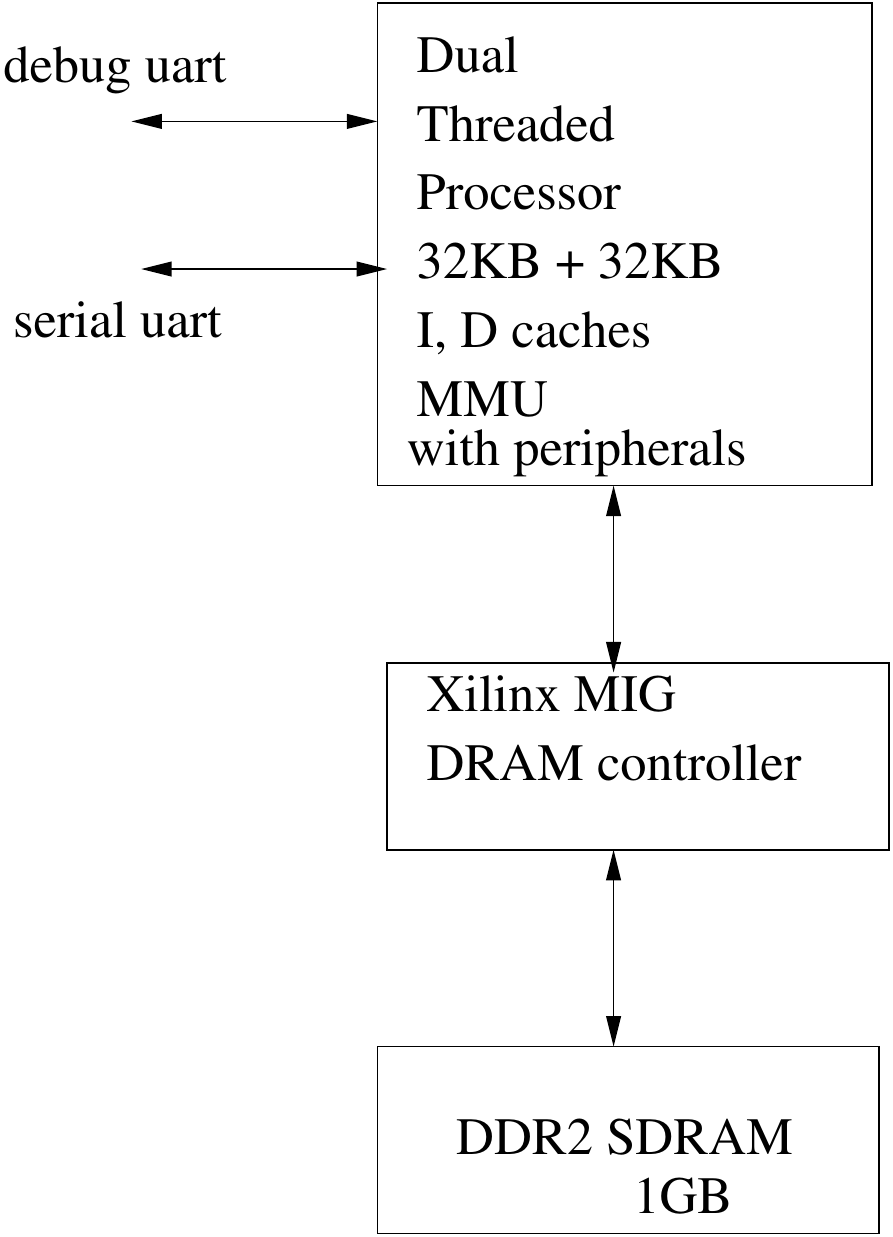}
  \caption{FPGA evaluation system for AJIT dual-threaded core}
  \label{fig:EvalSystem}
\end{figure}

A dual-threaded processor with 32KB each of instruction and data cache
was used in the evaluation.   The caches were 4-way set associative
and implemented a write-through-allocate protocol.  The implementation was done on
a KC705 FPGA card.  The system block diagram is shown in Figure \ref{fig:EvalSystem}.
The system is clocked at a frequency of 80MHz, and 
for this configuration, the cache miss penalty (for both instruction
and data caches) is 30 clock cycles.

In order to evaluate the performance of the dual threaded processor,
we identify the following scenarios in which the dual threaded processor
can be operated.
\begin{itemize}
\item Thread 1 is inactive and only thread 0 is used.  We call this the 
{\em thread 0 active, thread 1 inactive} scenario.
\item Thread 1 is spinning for work, and thread 0 is used for the actual
computation.  Here the activity of the spin loop on thread 1 will use
up some of the cache bandwidth.  It is interesting to know the
degree to which this affects the computation on thread 0.  We call it
the {\em thread 0 active, thread 1 spinning} scenario.
\item Thread 0 and thread 1 are used together for the cooperative computation,
using the side-kick programming model.  We call this the the
{\em Thread 0 active, thread 1 active} scenario.
\end{itemize}
While evaluating the performance, we measured the run times
of the three scenarios listed above, as well as the run time measured
on the single threaded AJIT processor.
The mearurements are summarized  in Figure \ref{fig:Speedups}.

\begin{figure}
  \centering
\begin{verbatim}
---------------------------------------------------------------------------------
Application      Single-threaded       Dual Threaded Scenarios           Speedup 
---------------------------------------------------------------------------------
                                    0 active    0 active     0 active    
                                    1 inactive  1 spinning   1 active
---------------------------------------------------------------------------------
Matrix-mult 
   seconds         1.6               1.6        1.64          0.85        1.88

Coremark           
  CM/MHz           2.42              2.42       2.30          4.36        1.80

ECG           
   seconds         0.59              0.59       --            0.33        1.79

FFT          
   seconds         0.167             0.167      0.172         0.094       1.78

Merge-sort     
   seconds         4.56              4.56       4.81          2.6         1.75

Dot-product 
   seconds         0.118             0.118      0.122         0.072       1.64

Bellman-Ford
   seconds         0.071             0.071      0.077         0.044       1.61

Daxpy
   seconds         0.133             0.133      0.149         0.111       1.2

Mutexes       
   seconds         1.025             1.025      1.05          0.872       1.18
  
Mem-copy     
   seconds         1.02              1.02       1.23          1.17        0.87
---------------------------------------------------------------------------------
\end{verbatim}
  \caption{Measured performance metrics for the four configurations.  Two copies of Coremark were run simultaneouly in the dual-threaded case.}
  \label{fig:Speedups}
\end{figure}

From the results, we observe that the performance of the
single thread processor is the same as that of the dual-thread
processor with only one thread active.  Further, we observe that
if both threads are active, with thread 1 spinning and thread 0 working,
there is a  small reduction in performance relative to the single
threaded case due to memory accesses generated by the spinning thread.   
The net increase in performance (last column of data relative
to first column of data) is greater than 1.6 for seven out of the ten examples
tried. 

In the ``Mem-copy''  case, a 32KB source array is being copied
to a 32KB destination array, and the miss-rate in the DCACHE is
87\%.   A data cache miss created by one thread can cause a stall in the second
thread if it wants to access the data cache.  

In the ``Mutexes''  case, there is very little
parallelism because only one of the threads will be in the critical
section at a time.   

Finally, in the DAXPY case, for $N$ dimensional vectors, there are
3N accesses to memory and only 2N computations.   The caches are
heavily utilized by both threads doing the computation and the
speedup obtained is modest.

\section{The ECG signal processing chain: a case study} \label{sec:ECG}

\begin{figure}
  \centering
  \includegraphics[width=12cm]{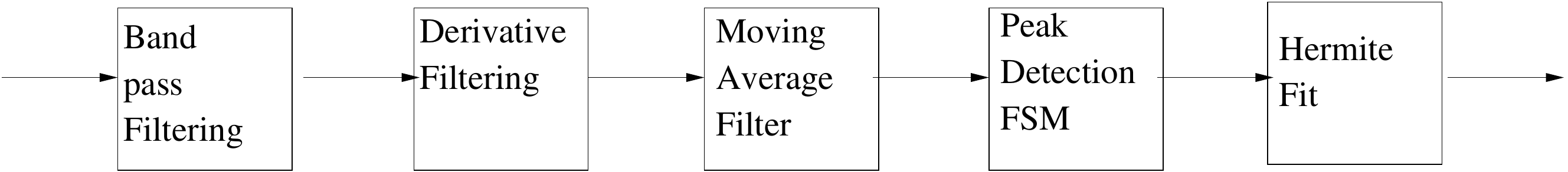}
  \caption{ECG signal processing chain}
  \label{fig:EcgChain}
\end{figure}

\begin{figure}
  \centering
  \includegraphics[width=12cm]{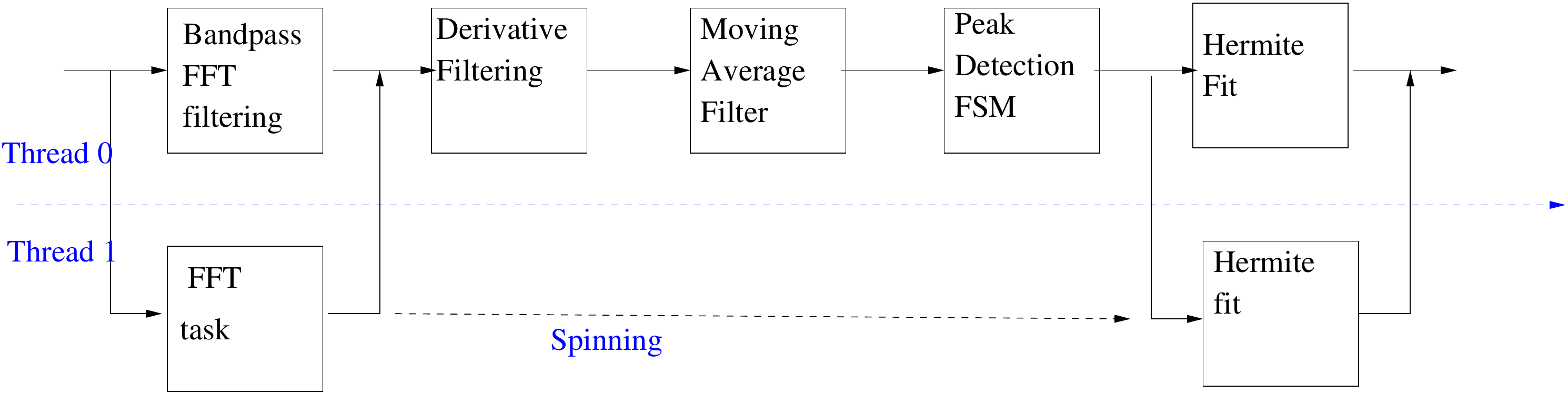}
  \caption{Parallelized ECG signal processing chain}
  \label{fig:EcgChainParallelized}
\end{figure}

We map the ECG signal processing application described in \cite{ref:EcgFpga},
which  implements the sequence of processing actions shown in Figure \ref{fig:EcgChain}.
The incoming signal is from an analog to digital converter and
consists of 12-bit data sampled at 360Hz.  The incoming data
is first band-pass filtered using a 256 point fast fourier transform,
which is parallelized across the two threads.  
The filtered signal is then analysed
to detect the centre of the heart beat (the QRS-peak).  This
analysis involves a derivative filter and a moving average
filter, followed by a finite state machine which makes a decision
on the peak at the centre of the heart beat.  This part of the
processing is inherently sequential and was not parallelized.   
Once a heart beat has been identified, a Hermite transform calculation is performed
to find the best representation of the heart beat as a sum of
Hermite polynomials.  This involves the calculation of a large number
of correlations  (dot-products), and is parallelized across the two threads.
The final parallelized implementation across the two threads is
shown in Figure \ref{fig:EcgChainParallelized}.

Of the total computing load in the signal processing chain, the
band pass filtering takes about 70\% of the time, and
the Hermite polynomial fitting takes almost 30\% of the time.
These two steps are parallelized across the two threads using
the side-kick programming model.  The observations are summarized
in Figure \ref{fig:EcgResults}.

\begin{figure}
  \centering
\begin{verbatim}
-----------------------------------------------------------------
         Total-time   band-pass-filtering  Hermite-fitting
           per beat       per beat            per beat
-----------------------------------------------------------------
single      35ms          25ms                9.6ms
thread

dual        19.5ms        13.4ms              5.8ms
thread
-----------------------------------------------------------------
\end{verbatim}
  \caption{Results for the parallelized ECG chain}
  \label{fig:EcgResults}
\end{figure}

The parallelization of the application across the two threads results
in a speedup of 1.79X.

\section{Summary}

We have described the architecture and an
evaluation of a dual-threaded shared cache
processor.  Using the excess available cache
bandwidth in a single threaded processor, we
add a second hardware thread which shares the
cache subsystem.  Two identical hardware threads are then
present in the processor, and these threads are
closely coupled via the cache subsystem.    
If it is not being used, it is possible to de-activate
one of the hardware threads in order to save power.   

The construction of the
dual threaded processor leverages an existing
implementation of a proven and verified 
single threaded processor.
The estimated area of the
dual threaded processor is 1.25X that of 
the single threaded one.

We use a cooperative
programming model to evaluate a set of applications on
the dual-threaded processor. 
In this model, hardware thread 0 is the main thread
and this thread can invoke a task on hardware thread 1
during the execution of the application program.  This
invocation is done through a shared memory region.  
The granularity of the invoked task can be as
small as 100 clock cycles due to the low invocation
overhead of 25 clock cycles.

We evaluated a range of applications on this
platform.  In applications that are possible to parallelize, 
the speedup observed relative to
a single-threaded processor is healthy, ranging
from 1.61X for the Bellman-Ford algorithm to 1.88X for matrix multiplication.  A more
complex and complete ECG signal chain application was also evaluated.
This too showed a speedup of 1.79X relative to the single-threaded processor.

When there is not much parallelism available, there 
is no benefit of the second thread, and it may be de-activated.
The ``Mutexes'' example illustrates that if the original application
itself is not parallelizable, the use of two threads does not offer any
significant benefit.

When the pressure on the cache from each of
the threads is high, or when the cache miss rate is high, 
the cache becomes the bottleneck and the performance of the dual
threaded processor is degraded.
The ``Mem-copy'' application exhibits a cache miss rate of 87\% 
and the  performance of the dual-threaded processor is
in fact worse than that of the
single-threaded one.  In such situations, just one of the hardware threads
can be used, and the idle one can be de-activated to reduce power dissipation.

In conclusion, the addition of the second thread is of benefit in
the following ways:  the area overhead is only 25\%, the performance improvement
on representative applications with internal parallelism 
is much larger than the increase in cost, and finally, if it is not advantageous to use both
threads, one of them can be shut off to save power.  
The dual threaded
processor with only one thread active has the same performance as the
original single threaded processor.

Other programming models remain to be explored.  Further,
in an embedded systems scenario, the impact of the
dual threaded processor in real-time applications merits
deeper analysis.   In particular the increase in interrupt handling
capacity in the dual-threaded processor can yield 
benefits to the embedded system designer.  

\bibliography{refs}

\end{document}